\pdfoutput=1
\documentclass{sigchi}

\toappear {
{* This is a pre-print version of the following article:}\\
Maria Karyda, Merja Ry\"{o}ppy, Jacob Buur and Andr\'{e}s Lucero.\\ Imagining Data-Objects for Reflective Self-Tracking \\
\textit{Proceedings of CHI '20}\\
\url{http://dx.doi.org/10.1145/3313831.3376844}}

\usepackage{xcolor}
\usepackage[framemethod=tikz]{mdframed}

\definecolor{cccolor}{rgb}{.67,.7,.67}

\CopyrightYear{2020}
\setcopyright{acmlicensed}
\doi{https://doi.org/10.1145/3313831.XXXXXXX}
\isbn{978-1-4503-6708-0/20/04}
\conferenceinfo{CHI'20,}{April  25--30, 2020, Honolulu, HI, USA}
\acmPrice{\$15.00}

\usepackage{balance}       
\usepackage{graphics}      
\usepackage[T1]{fontenc}   
\usepackage{txfonts}
\usepackage{mathptmx}
\usepackage[pdflang={en-US},pdftex]{hyperref}
\usepackage{color}
\usepackage{booktabs}
\usepackage{textcomp}

\usepackage{microtype}        
\usepackage{ccicons}          

\def\plaintitle{Imagining Data-Objects for Reflective Self-Tracking*}

\def\emptyauthor{}
\def\plainkeywords{Authors' choice; of terms; separated; by
  semicolons; include commas, within terms only; this section is required.}

\makeatletter
\def\url@leostyle{%
  \@ifundefined{selectfont}{
    \def\UrlFont{\sf}
  }{
    \def\UrlFont{\small\bf\ttfamily}
  }}
\makeatother
\def\pprw{8.5in}
\def\pprh{11in}

\definecolor{linkColor}{RGB}{6,125,233}
\hypersetup{%
  pdftitle={\plaintitle},
  pdfauthor={\emptyauthor},
  pdfkeywords={\plainkeywords},
  pdfdisplaydoctitle=true, 
  bookmarksnumbered,
  pdfstartview={FitH},
  colorlinks,
  citecolor=black,
  filecolor=black,
  linkcolor=black,
  urlcolor=linkColor,
  breaklinks=true,
  hypertexnames=false
}

\usepackage{scrextend}
\usepackage{cuted}
\usepackage{capt-of}

\begin{document}

\title{\plaintitle}

\numberofauthors{4}
\author{
 Maria Karyda\textsuperscript{1}, Merja Ry\"{o}ppy\textsuperscript{2}, Jacob Buur\textsuperscript{2}, Andr\'{e}s Lucero\textsuperscript{1} \\
 \\
  \affaddr{\textsuperscript{1}Aalto University, Espoo, Finland}\\
  \affaddr{\textsuperscript{2}University of Southern Denmark, Kolding, Denmark}\\
  \email{maria.karyda@aalto.fi, \{merja, buur\}@sdu.dk, lucero@acm.org}
  }

\maketitle

\begin{strip}
\centering
\includegraphics[width=\textwidth]{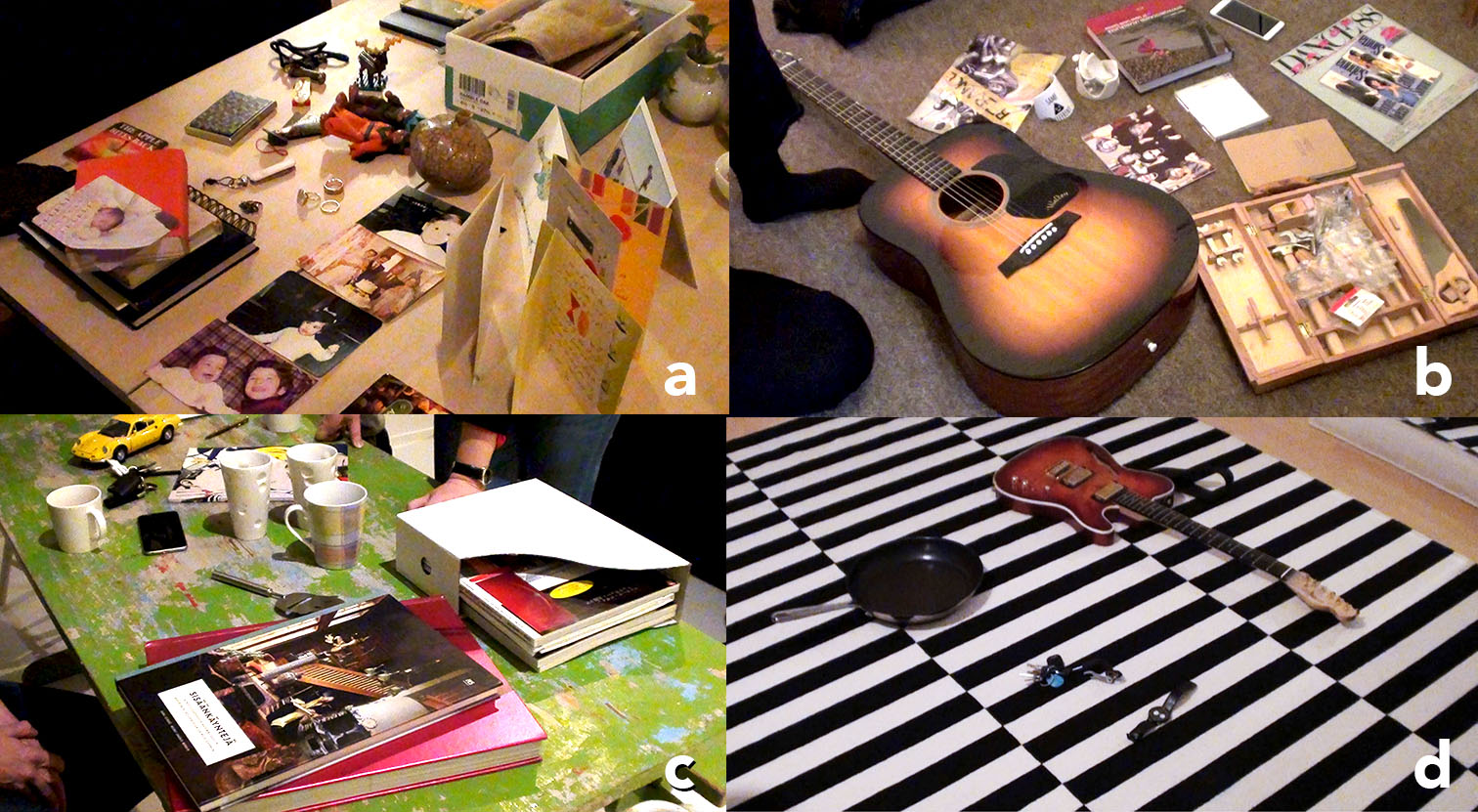}
\captionof{figure}{Four collections (a-d) of personal objects presented at four participants' households. Each participant chose and used their own objects to imagine novel data-object combinations based on their self-tracking practices.}

\label{fig:collages}
\end{strip}

\begin{abstract}
While self-tracking data is typically captured real-time in a lived experience, the data is often stored in a manner detached from the context where it belongs. Research has shown that there is a potential to enhance people's lived experiences with data-objects (artifacts representing contextually relevant data), for individual and collective reflections through a physical portrayal of data. This paper expands that research by studying how to design contextually relevant data-objects based on people's needs. We conducted a participatory research project with five households using object theater as a core method to encourage participants to speculate upon combinations of meaningful objects and personal data archives. In this paper, we detail three aspects that seem relevant for designing data-objects: social sharing, contextual ambiguity and interaction with the body. We show how an experience-centric view on data-objects can contribute with the contextual, social and bodily interplay between people, data and objects.

\end{abstract}




\section{Introduction}
Self-tracking is argued to be an exploratory process that helps people to gain control over their lives through self-reflections or, put differently, to \textit{``achieve self-knowledge through numbers''} \cite{Choe:2014:UQP:2556288.2557372,Li:2011:UMD:2030112.2030166,williams2015anxious}. Previous research has shown that people track for different reasons and have developed a variety of ways to suit the specific needs for capturing data. Sleep, exercise, sex, food, mood, location, alertness, productivity and well-being are some of the things people choose to track \cite{wolf2010data}. This tracking is done by using a variety of tracking systems \cite{Lupton:2014:SCT:2686612.2686623}. Tracking devices frequently come in a form of wearable technology to capture physiological data or mobile phone applications that allow manual tracking. Representations of data from the current devices come as a screen-based numerical output, visualized in graphs. However, often people create their own data visualizations \cite{pousman2007casual} or use paper journals to track their data \cite{Elsden2017DesigningInformatics,Ayobi:2018:FMS:3173574.3173602}. Besides personal visualizations, people's self-driven need for exploration has been addressed also through physical representations of data \cite{Barrass2012}. Data physicalizations such as physical data sculptures or sonifiations of data \cite{Barrass2012,moere2008beyond,Sosa2018,Karyda:2017:CCI:3098279.3119927}, invite people to reflect on data in alternative ways. 

This exploratory engagement happens at a distance from the actual living experience where the data was captured. For instance, people's spreadsheets may allow them to engage in self-reflections, but only in retrospect. \textit{Data-objects} \cite{Sosa2018,Zhu2015Data-objects:Interfaces} are artifacts that represent contextually-relevant data, offering the opportunity for exploratory engagements in the context where data was captured. As an approach to self-tracking and data physicalization, the concept of data-objects opens up a whole gamut of opportunities and challenges. Some of them have already been articulated by previous work \cite{Karyda:2017:CCI:3098279.3119927,Sosa2018,Zhu2015Data-objects:Interfaces}, yet those challenges do not adequately show people's needs and how data-objects can be implemented into people's everyday life. Through our study we offer a reflective view on the three-way interplay between people, data, and artifacts, which opens the ground for further explorations of data-objects. 

Since data-objects are situated in lived experiences, we conducted our study in six participants' homes. This enabled us to experience different ways in which people engage in tracking and how they interact with their data. This understanding enabled us to build upon our participants' self-tracking practices to collectively imagine data-objects. We challenged the participants to show their most cherished artifacts and used them as object probes \cite{DeLeon2005} to gain first-hand insights into people's social, material and body-centric relationships \cite{woodward2016object,Moller:2018:PAA:3173574.3174133} to objects and tracking. We based the probing session on object theater exercises \cite{myatt2012frozen,ryoppy2018object}, where the participants imagined speculative combinations of objects and data (e.g., a guitar that captures and plays back your sleep data), mildly related to their tracking habits.

We pose the following research questions: 1) What can we learn from people's current self-tracking practices in informing data-objects? 2) How can we encourage speculative combinations between everyday objects and data? 3) What are the implications for designing data-objects for situated data representation? Our results allow us to better define the nature of data-objects and expand on the design opportunities and challenges with those systems. The contribution of our work is two-fold: a) we suggest new methods to investigate how people understand and reflect on data, and b) we introduce directions on how to design data-objects from an experience-centric point of view.

In the following sections, we articulate the current landscape in personal informatics and data physicalization in an effort to contribute to the current understanding of data-objects. We then highlight that a consideration has to be taken towards designing tracking tools that provide people with the opportunity to capture, interpret, reflect and discuss their data onsite. Next, we provide an overview on the origins of our methodology.  

\section{An Experience-Centric Approach to Data}
The current turn in HCI towards an exploratory, experience-centric approach \cite{McCarthy:2004:TE:1015530.1015549} offers new ways to understand how people engage with their data. Rooksby et al. suggest the term \textit{lived informatics} to describe how people use information and find its meaning in their daily lives \cite{Rooksby:2014:PTL:2556288.2557039}. The authors report on how people use trackers to document different aspects of their lives and emphasize that interweaving trackers is not a rational process but a way to explore data. As an experience-centric approach, they draw on McCarthy and Wright \cite{McCarthy:2004:TE:1015530.1015549} to dispense with the idea that personal informatics tools are primarily there to objectively reveal data \cite{li2010stage}. Elsden et al. discuss how people make sense of their data retrospectively and map out design challenges around the long-term use of personal informatics \cite{Elsden2016AInformatics}. Both support the critical turn towards experience, but while Rooksby et al. \cite{Rooksby:2014:PTL:2556288.2557039} argue that people \textit{``choose, use, interweave and abandon devices''} (hence, it is meaningless attempting to design for long-term use), Elsden et al. \cite{Elsden2016AInformatics} propose that reminiscing of historical data can support the long-term use of personal informatics in a \textit{fertile} design space. Time-Turner \cite{Singhal2017Time-Turner:Home} is an example which supports this kind of reminiscing, allowing families to see and reflect on past family members' data through a set of an everyday object (i.e., coasters). The family can view pictures and videos in a way that data becomes part of the mundane and reinforces everyday remembering of data. 

Personal (physiological) data has its own materiality \cite{lupton2016personal}, it may be characterized as meaningful digital possessions \cite{Elsden2016AInformatics} and as boundary objects, when represented through a data visualization or physicalization \cite{Buur2018Physicalizations,mortier2014human}. The concept of Data Physicalization \cite{Jansen:2015:OCD:2702123.2702180} excellently illustrates a dialogue between the digital and physical manifestations data might take. Data Physicalization investigates (personal) data by inviting people to experience information in visual, haptic and in some cases sonic ways \cite{Barrass2012,Nissen:2015:DDF:2702123.2702245}. Examples include the DNA ring \cite{Rezaeian:2014:DTD:2636240.2636869}, Barrass's singing bowl representing blood pressure \cite{Barrass2012}, and Frick's sleeping patterns \cite{Frick}. Common among those examples is that personal data is self-organized in unique ways to provoke meaningful reflections. 

Data-objects \cite{Sosa2018} are physical representations of data at the intersection of data physicalization \cite{Jansen:2015:OCD:2702123.2702180}
and industrial design. Data-objects provide a platform for data curation and experimentation through an experience-centric approach \cite{McCarthy:2004:TE:1015530.1015549} to data. In their material form data is embodied. Similar to everyday objects which are situated within certain activities, data-objects are contextually appropriate and able to be linked to people's lived experiences. Current examples of data-objects demonstrate static representations of everyday objects, which do not allow for any type of interaction with a person apart from the haptic \cite{Zhu2015Data-objects:Interfaces}. 

For instance, in Zhu et al.'s \cite{Zhu2015Data-objects:Interfaces} personal data is represented physically on the surfaces of common objects, like a coffee mug that can be carried around. Karyda's \cite{Karyda:2017:CCI:3098279.3119927} work instead demonstrates examples of modified artefacts that allow people to interact with their data in novel ways. In her work, everyday objects (like a data-modified guitar plectrum) allow people to revisit certain datasets touching upon the aspect of reminiscing, described by Elsden et al. \cite{Elsden2016AInformatics}. These examples illustrate how digital data may potentially be part of people's physical and personal surroundings facilitating information to people not only in static but also in dynamic ways. 

On a different note, Sosa et al. \cite{Sosa2018} introduce the idea of Data-Objects and Design Activism. In that example, data-objects enable people to make sense of information that is useful to enhance society. Here, physical representations of data become relevant and important for the general public addressing a contentious call for change. Taylor et al. \cite{taylor2015data} argue that data is bound up with the place both in physical and social geography. The latter describes how data is topical in a communal level that goes beyond the individual. Both \cite{Sosa2018} and \cite{taylor2015data} illustrate how data-objects can also go beyond the individual towards objects that combine datasets of multiple people or are significant to many. In our work, envisioning data-object combinations allows tracking and data to be integrated in everyday interactions considering the place and the people. In the following section, we dig deeper into the role of objects in everyday life, which was an inseparable part of our methodology.

\begin{figure*}
\centering
\includegraphics[width=\textwidth]{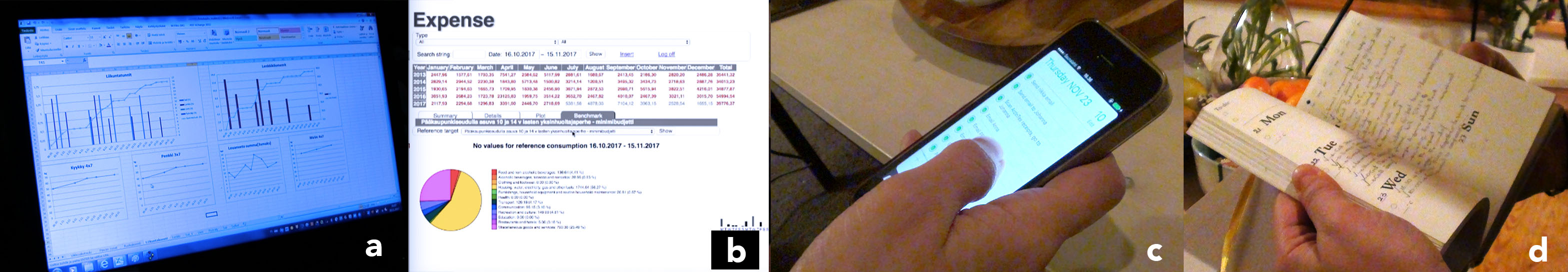}
\caption{During our visits at the participants' homes we found several self-created tracking tools. From left to right: a) Sport digital spreadsheets, b) Expenses in an online system, c) Task lists in the phone, d) Task lists in the notebook.}
\vspace{-1em}
\label{fig:figure1}
\end{figure*}

\section{Object Studies}
People can often recall meaningful moments by touching or seeing an object. Ghosh quotes Carrignon, an object theater artist, when claiming: \textit{``Objects are memory boxes. They trap within themselves individual memories and collective memories''} \cite{Ghosh}. Digital objects, such as trackers, allow people to access another kind of memory through technology. This complex relationship between people, artifacts and memory calls for a new understanding. This section describes what inspired our methodology and explains why we used people's personal objects in envisioning future data systems. We present three major approaches: 1) Social, 2) Material and Body-centric; and 3) Transformative Relationship with objects, which all draw from different methods. 

\subsection{Social Relationship with Personal Objects}

Research on personal objects and how people value their possessions, has a long tradition in design anthropology. Latour's \cite{latour1987science} actor-network theory considers objects and people constantly shifting in networks of relationship. Miller \cite{miller2008comfort} is committed to ethnographic exploration of how people possess everyday objects and connect artefacts with human experiences of loss, longing, grief, death, attachment and love. Objects are seen to shape human experiences, in particular by addressing the aspects of symbolic attachment and significance that objects gain in social relationships. DeLeon and Cohen \cite{DeLeon2005} have developed an ethnographic interview method, called \textit{object probe} that helps the participant recall events by interacting with objects. They have witnessed how old objects, such as photographs and musical instruments, evoke important memories of people, places, and events of a person's life. A physical object can help a person share memories of a specific era in their life, such as childhood, university studies, or the birth of a first child \cite{DeLeon2005}. We used object probes by challenging people to show their most cherished artifacts, which gave us first-hand insights into people's social relationship with and through personal objects.

\subsection{Material and Body-Centric Relationship with Objects}

While DeLeon and Cohen \cite{DeLeon2005} highlight the importance of objects in sharing personal memories, Woodward \cite{woodward2016object} introduces the method of \textit{object interview}s to explore how people articulate their material and lived experiences. In this way the researcher can gain empathy that extends beyond \textit{``what is being said,''} and includes \textit{``how it is being said,''} gestures, objects, material qualities and colors. By presenting the example of a pair of jeans, Woodward illustrates how \textit{``the material properties of things are central to understanding the sensual, tactile, material and embodied ways in which social lives are lived and experienced''} \cite{woodward2016object}. According to Moeller \cite{Moller:2018:PAA:3173574.3174133} the body-centric relationship to wearable health products is important as it offers insights into personal style and choice of aesthetics, articulating the cultural fit of a design object to a particular individual. In our study, thinking about personal data archives as \textit{wardrobes} and tracking devices as accessories was particularly useful while digging into material and body-centric relationships that our participants have with their tracking tools and personal data archives.
\subsection{Transformative Relationship with Objects}

In the current approaches of object probes \cite{DeLeon2005}, object-interviews \cite{nordstrom2013object, woodward2016object}, accessory \cite{Moller:2018:PAA:3173574.3174133} and wardrobe studies \cite{Klepp2014}, artifacts are used by researchers to learn about people's existing social, material and body-centric relationship to sentimental objects and materiality. While the approaches bring a nuanced understanding to personal possessions, it is yet unclear how to benefit from that knowledge in envisioning future digital objects. We extend these approaches by engaging people in imagining future interactions with objects through \textit{object theater}, a process of telling stories through and with objects \cite{myatt2012frozen, margolies2016props}. In object theater, the performer manipulates an object to tell a story for an audience, and by doing so challenges the use of the artifact by transforming the relationship between people and objects. The most well-known (and highly debated) form of object theater employs objects as living puppets. In other forms, the performer positions the objects in syntactic relationship to other objects and focuses on certain attributes of the object. For example, in the hands of a performer a one-meter tape measure can turn into: a story of a person trying to lose weight, a timeline of one year, or a living puppet (e.g., a snake). 

Object theater exercises help explore mundane objects and their properties to tell stories. According to Margolies: \textit{``when objects arrive in the workshop of artists, they are already charged, by virtue of the wear of the material and their former life (...) What is required then, is to recharge them, that is to say, to make them visible, to bring out a certain expression, a sign, a metaphor''} \cite{margolies2016props}. For object theater teacher Rene Baker \cite{Baker}, objects are cultural, and people need to first work with those cultural connections before starting to develop a story. A metaphor arises from the cultural heritage that people have with objects. The object brings to the performance \textit{the charge} of their previous history and \textit{the recharge} of new metaphors to be interpreted by the spectator. 

In traditional object theater \cite{myatt2012frozen} the performer tells stories to the audience using \textit{ready-made objects} on stage. Our work introduces object theater exercises to interaction design and participatory design traditions, for examining the relationships between the object, the (design) researcher and the (design) participant. The research builds on previous work by Buur and Friis \cite{44b44fd5eaee419ca9729b85acb875ef}, but differently from their four object theater perspectives developed with design students in a studio, we show how a novel object theater approach can be extended to work with people in the field. With object theater in the field studies, both the researcher and the participant are immersed in the context and personal space of the participant, making use of personal objects selected and introduced by the participant. This expands Ry\"{o}ppy et al.'s \cite{ryoppy2018object,ryoppy2017exploring} research on the use of object theater in field interviews, as it borrows from theater improvisation, probes and generative design methods, which mainly make use of ready-made objects selected and brought by the researchers. We used object theater techniques to recharge participants' personal objects with the attributes of their tracking practices to inspire stories of novel ways of combining personal data with everyday objects. The participants' stories about personal objects that carry their own history inspired unconventional combinations of novel data-objects. Imagining unique data-objects based on experiences with existing sentimental objects has practical significance for designing novel artifacts. The process both reveals past experiences bound to the objects and serves as a trigger for imagining future aspirations.

\section{Methodology}
Self-tracking and personal informatics has been researched with questionnaires, survey and qualitative interviews \cite{Elsden2016AInformatics, li2010stage, Li:2011:UMD:2030112.2030166, Rooksby:2014:PTL:2556288.2557039}. To study the multitude of digital, visual and tangible forms that data takes, we adopted new methods inspired by object theater and followed a participatory design approach. 

\subsection{Study Design}
Our participants were recruited over a period of three weeks via a short online questionnaire with the main criteria of having 1) an interest in, and 2) prior experience with self tracking. Interested people replied to three questions: \textit{What do you use to track your data? What kind of data do you track? Why are you interested to join the study?} Our main challenge was to find participants, who would allow us into their homes and be committed throughout the study period of six months. Hence, seven initial participants were recruited in the following ways: one through posters at the University, two through the online forum MyData\footnote{https://mydata.org/finland/}, three through personal networks (i.e., Instagram, Facebook, WhatsApp) and professional channels (i.e., LinkedIn), and one through snowball sampling. These seven respondents completed the initial questionnaire and one later dropped out. The six participants (2 females, 4 males) lived in the metropolitan Helsinki area. While challenging, our recruitment process provided enough participants to conduct an in-depth participatory experiment capturing detailed idiosyncratic accounts of people's cherished possessions.

Following the recruitment process, our study proceeded in two parts, from gathering general information about tools and practices for self-tracking to specific stories about objects perceived as personal. 

To surface values that cannot be expressed through talking, we visited our participants' homes. All the field visits were video recorded and later transcribed. With each participant, we conducted an exploratory object interview together with an object theater exercise, which lasted 60-90 minutes per participant. We combined traits of object probe \cite{DeLeon2005}, wardrobe studies \cite{Klepp2014}, object-interviews \cite{nordstrom2013object,woodward2016object} and accessory approach \cite{Moller:2018:PAA:3173574.3174133} to scaffold participants in explaining their practices and preferences. 

The visits enabled the participants to show their digital trackers and self-created tracking tools (as found out during our visits), demonstrating what they do with the trackers and how they interact with their collected data (Figure \ref{fig:figure1}). Building upon the self-tracking practices, we challenged our participants to envision data-object combinations. In bringing different artifacts into our conversation, we asked questions such as \textit{Can you describe an important object from each decade of your life?}, as well as, personal objects e.g., \textit{Can you show me where you store this object? Can you show me how you use it?} The objects presented by the participants were brought together to form an assembly (Figure \ref{fig:collages}). The personal objects varied from ordinary everyday objects, such as keys, musical instruments, sports equipment and books, to sentimental artefacts, such as gifts, heirlooms and photo albums. Every time an object would be introduced, a story was brought up about the origin of that object. These stories led to discussing people's memories and relations to others, around the object at hand. In the object theater exercise the participants were encouraged to envision new combinations of their personal objects and tracking, e.g., \textit{How would you register your data on X object?} or \textit{How would you read the data from the object?} These kinds of insights would have been impossible to surface through a standard interview, as the tangibility of the objects simultaneously enabled material expression, unsettling of pre-existing interpretations and reflective evaluation of participants preferences. 

\subsection{Data}
Table \ref{tab:Demographics} provides an overview of the participants and their motivations to join the study. Table \ref{tab:Table2} shows what the participants track, which systems they use and which personal objects they presented as valuable to them during the interviews. The table includes also a summary of data-object ideas that the participants came up with. Names are pseudonymized.

\begin{table}
\centering
\small
\begin{tabular}{ p{0.6cm}| p{0.2cm} p{6cm} } 
\textbf{Name} & \textbf{Age} & \textbf{Motivation} \\
\hline 
Hans & 30 & Interested in understanding personal objects as data gathering devices. He finds the long-term use of multiple applications and devices at the same time challenging. Alters between periods of intensive tracking and not tracking at all. \\
\hline
Simon & 32 & Interested in the project, as he said: \textit{``I could sometimes use a better tool for tracking.''} He alters between periods of intensive tracking and not tracking at all.\\ 
\hline 
Max & 37 & Motivated to support research activity close to his interests. What he enjoys in tracking is how things change and develop over time.\\ 
\hline 
Anna & 24 & Joined the study out of curiosity. She has only recently started to track few aspects in her life.\\ 
\hline 
Scott & 41 & Wants to learn about novel ideas. He is extremely interested in data and tracking, and has done self-tracking over 10 years.\\
\hline 
Olivia & 34 & Interested in recognizing her own patterns and habits more consciously. She has been tracking her daily activities for years.\\
\hline 
\end{tabular}
\vspace{0em}
\caption{Demographics of all six participants.}
\vspace{-1em}
\label{tab:Demographics}
\end{table}

\subsection{Analysis}
For our analysis, the data consisted of videos from the field, transcripts of the six visits, 68 photographs of objects and six photographs of object assemblies. The photographs acted as reminders of the visits during the sense-making process. Based on a close reading of the transcripts, the two first authors identified how the idiosyncratic traits of each participant were reflected through their objects and tracking habits. 

The analysis of the data-objects drew from Gaver and Bower's \cite{Gaver:2012:AP:2212877.2212889} annotated portfolios. We extracted the data-object ideas from the transcripts and used a similar labeling process, as if the ideas were physical prototypes. Clustering and analysing the data-object ideas based on interaction styles, the type of data and the motivation of the participants gave us seven themes:  \textit{self, others, close to the body, away from the body, representation, tracking} and \textit{connectedness to the data}. Then, a synthesis of the themes gave us the three categories of the results: representation, self and
others formed \textit{`Social Sharing'}; connectedness to the data formed \textit{`Contextual Ambiguity'}; and
closeness to the body tracking and representation formed \textit{`Interacting with the Body.'}

\section{RESULTS}

In the following section, we present the tracking habits of our participants and the combinations they envisioned one by one, to show who the participants are and why they proposed such data-object combinations. We then present the results of our analysis on the data-object combinations.

\subsection{On Tracking}

\textit{Hans} started early to track his soccer practice on digital spreadsheets. \textit{``I've been making Excel sheets since primary school, like since I was 10 years old or something. I did it just as a hobby. I put my soccer practices on Excel sheets and then I counted how many hours I've played football and how many bounces I have been able to do.''} This practice of self-tracking continues until today, altering between periods of intensive tracking and no tracking at all. He has been using other ways to keep track of his data such as on Huawei Health\footnote{https://play.google.com/store/apps/details?id=com.huawei.health\&hl=en\label{Huawei}} application, a smartwatch and notebooks. The main data Hans captures is sports exercise and food consumption. He manually inputs his total hours for exercising, running distance and duration, gym exercises and amount of weights to the digital spreadsheet and creates his own visualizations to combine those data points he is mostly interested in, presented in Figure \ref{fig:figure1}a. 

\begin{figure}[t!]
\centering
\includegraphics[width=8.4cm,height=8cm]{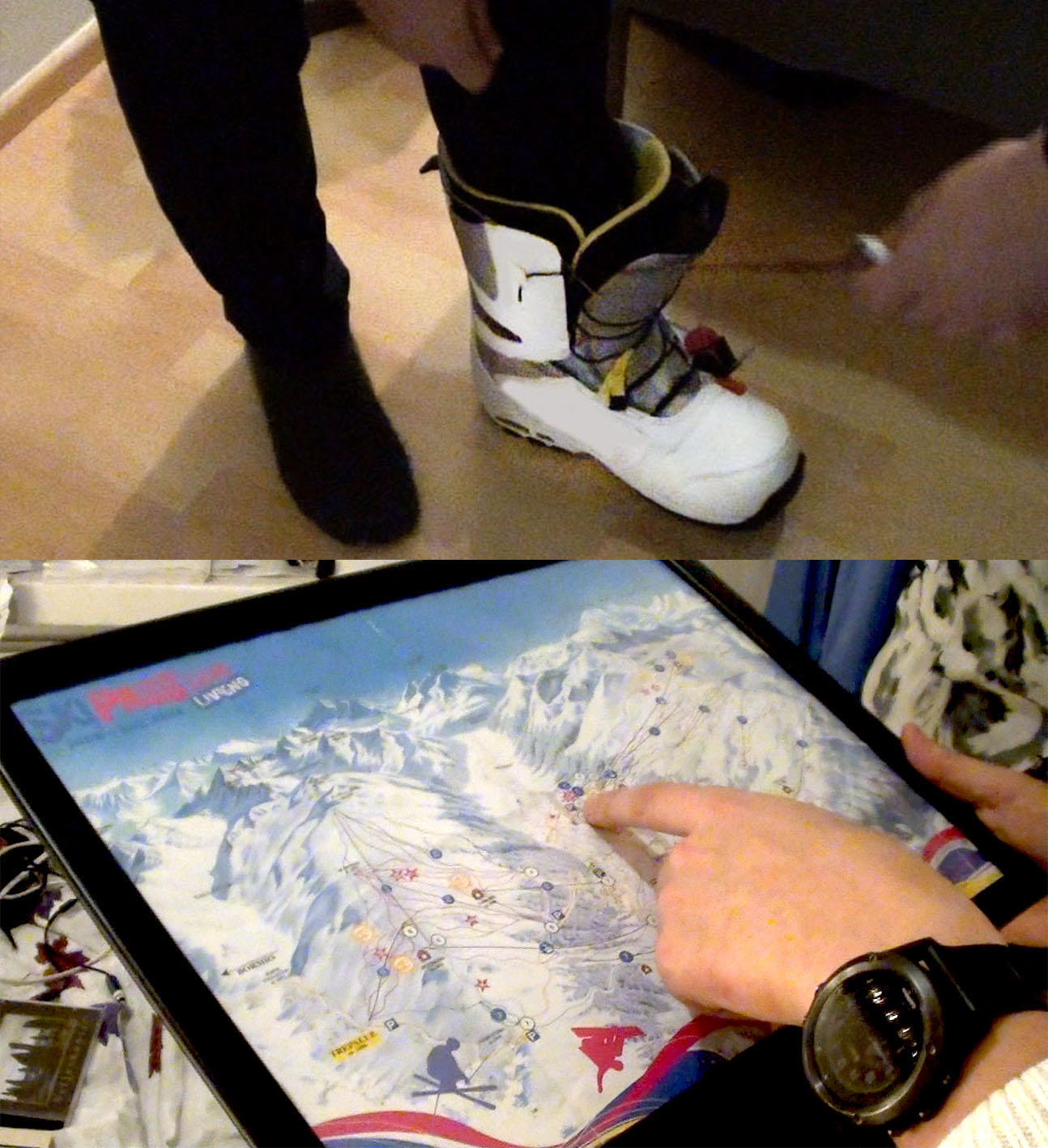}
\vspace{0em}
\caption{Hans demonstrates how he would use his skiing boots and a map to track himself.}
\vspace{-1em}
\label{fig:boots}
\end{figure}

\textit{``I like to make those Excel spreadsheets, [the reason] is that I like to make my own kinds of, well not my own kinds of, but visualize those things that I want to be visualized. And group the things together that I feel the need to be grouped together.''} He has created another spreadsheet for food, because the existing phone applications for food tracking were providing too many options. \textit{``There were millions of different foods in them. I have somewhere around 30 different ingredients here, so it's easier for me to use it. It's those things that I eat. I know I eat.''} In the future, Hans would be interested to track more data, such as, speed, heartbeat or sleeping data in order to improve his performance. 

When envisioning future tracking objects, Hans was concerned that the tracking of a sports activity did not start at the right moment. While interacting with his snowboard boots, he made an action of tightening the laces which would start the tracking. He suggested objects that would automatically trigger logging based on an action. The tracker would preferably follow a person's actions and adapt to his lifestyle. For instance, he envisioned a combination of a skiing map, skiing boots and a snowboard that would track his skiing routes (Figure \ref{fig:boots}). The act of tightening his snowboard boots would start the tracking system while a folded paper map in his pocket would show the route.  Unfolding the map would stop tracking. Also, he thought of the skiing boots as an element that could track both his heartbeat data and the distance he covered while skiing. This suggests using a single device to track multiple datasets similarly to smart watches that can track multiple datasets e.g., sleep data, hear rate and more. The same vision of using a single device to track different datasets was also seen by Simon and Scott and it will be described below. Hans also imagined a system that would track his sleep by combining a book and his bed sheets. Thus, when the book is placed down the bed sheets would start tracking data. 

\textit{Simon} and Hans are roommates, hence they are to some extent influencing each other's habits. Simon's current tracking tools are Huawei Health\footref{Huawei} application, a beer evaluation application Untapped\footnote{https://untappd.com/}, a notebook and digital spreadsheets. He mainly uses the mobile application when hiking and the notebook to register how many weights he lifts at the gym. Simon tracks his food consumption in a similar way as Hans. Other elements he chooses to track are: walking and jogging distances, calories consumption, time at the gym and different beers drank. Something that became clear during our interview was that Simon performs tracking for short periods of time, three months, and then he moves to something new. He did not seem to have a way of combining and curating data but the different data sets remained separate on different platforms. 

Simon was keen on tracking his sleep data and imagined three different data-objects for that purpose: a guitar, a soft-toy, and a watch. All of them would record his sleep and play it back in the morning through sound. In addition, he thought of his keys as an object to track his heartbeat data and distance every day. Lastly, he imagined his frying pan being able to capture the calories of a meal and present them to him through different colors which would appear on the surface of the pan. For example, if the food would contain a lot of fat the pan would turn red, while low calories food would turn the pan green. 

\textit{Max} uses the Strava\footnote{https://www.strava.com/mobile} application for cycling, digital spreadsheets, his car's computer, and personal notes to capture data. The cycling application specifically is attractive to him because, apart from showing his routes and kilometers, it also allows him to compete with other cyclists. The winner of each route receives a badge for being the king of the mountain. Out of curiosity, Max had tracked the exact hours he spent on writing his doctoral dissertation on a digital spreadsheet. This is something he still does with his work hours and activities. Other data he used to track was activities, exercising, his family's food consumption, and driving data. 

Max imagined the future tracking of his cycling routes. Tracking would start by unlocking the bike and it would stop when the bike would be locked again. He envisioned personal mugs, which would indicate with different colors who is \textit{the king of the mountain}. For instance, Max imagined a cycling magazine that would combine datasets from different people every month: \textit{``A section introducing the route of the moth, this was discovered by Max.''} This would be a system that compares tracking routes from different people. On a different note, Max showed us a photo of his old jeans and imagined a trousers label that would show how fast they wear out throughout time.

\textit{Anna} employs analogue ways of tracking using notebooks. She is a dance teacher and uses a name list to check her students' attendance. While this is not self-tracking, the datasets she registers are relevant to her: \textit{``That's very unorganized, it's the back page of my notebook with my plans for the classes I just put like [Signs a check mark in air] if they're there. Well, in another one of the classes where I teach, every few weeks I have to transfer it to their sheets.''} For Anna the notebook is easy to use compared to a mobile application since it is physical, \textit{``a quick way to do it,''} which will not let her down. Another habit she has started was to list on her mobile phone the movies she has watched. She likes to remember which movies she has already seen, and in the future she imagines to share that list with friends who ask her for movie recommendations. In terms of digital tracking, Anna uses the Eve\footnote{https://play.google.com/store/apps/details?id=com.glow.android.eve\&hl=en} application to track her menstruation cycle and finds it more convenient than using a physical calendar, because it gives extra information of the period cycle and \textit{``I remembered to do it.''}

\begin{figure}[t!]
\centering
\includegraphics[width=8.4cm,height=6cm]{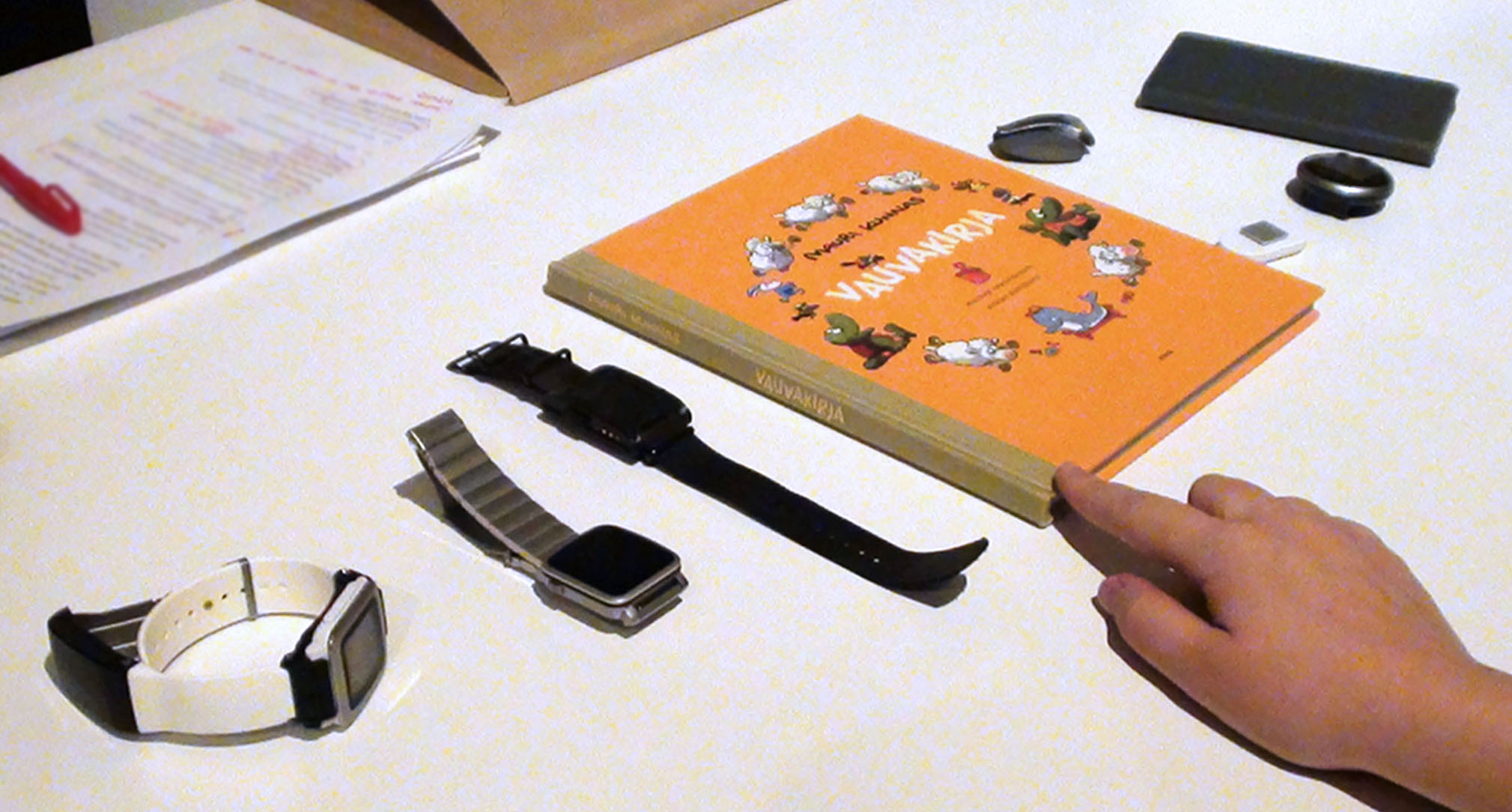}
\vspace{0em}
\caption{Scott's tracking devices, and a family album that he imagined as a data-object which combines data from all the family members.}
\vspace{0em}
\label{fig:Scott}
\end{figure}

In terms of the data-object combinations, she imagined a mug, an LP player and an album that would show the movies she has watched in digital and physical ways. She also imagined that she could use guitar notes to represent the level of pain during the menstruation cycle. She made a similar connection to represent her menstruation cycle and the pain by hammering nails on a surface resembling an art piece. Lastly, she came up with an idea of a photo-book that combines photos, videos and songs from different eras of her life.

\textit{Scott} has been self-tracking for almost 20 years. He owns seven tracking devices (Figure \ref{fig:Scott}) and uses all of them on a daily basis. Lumo Lift\footnote{https://www.lumobodytech.com/lumo-lift/} is a wearable device for tracking and maintaining a good body posture while sitting or working. Attached to his clothing with a detachable magnet, it uses an accelerometer to detect a person's posture and vibrates when the posture is bad for too long. A second device is a \textit{breath stone} tracker to monitor his breathing and how it correlates with the other data he collects. A Pedometer tracks his steps and compares the collected data with the data of his pebble watch and a mobile phone health app. When asked if Scott looks at the data from different devices every day, he stated: \textit{``I used to, but when you do that for a month you're not learning anything new just by looking at [it] every day, then you start checking it day to day and then weekly and monthly.''} He owns two earlier versions of the smartwatch: the kickstarter version that he received from backing the project, plus a developer's version. In addition to being inherently interested in tracking with wearable devices, Scott has created a web system for noting down his family's expenses (Figure \ref{fig:figure1}b). The system generates visualizations of the registered expenses in pie charts. Through this application Scott can cluster purchase data based on each person in the household, date, and time. The latest addition to his system is allowing online data retrieval, to compare his family's consumption to a general average.

\begin{figure}[t!]
\centering
\includegraphics[width=8cm,height=4cm]{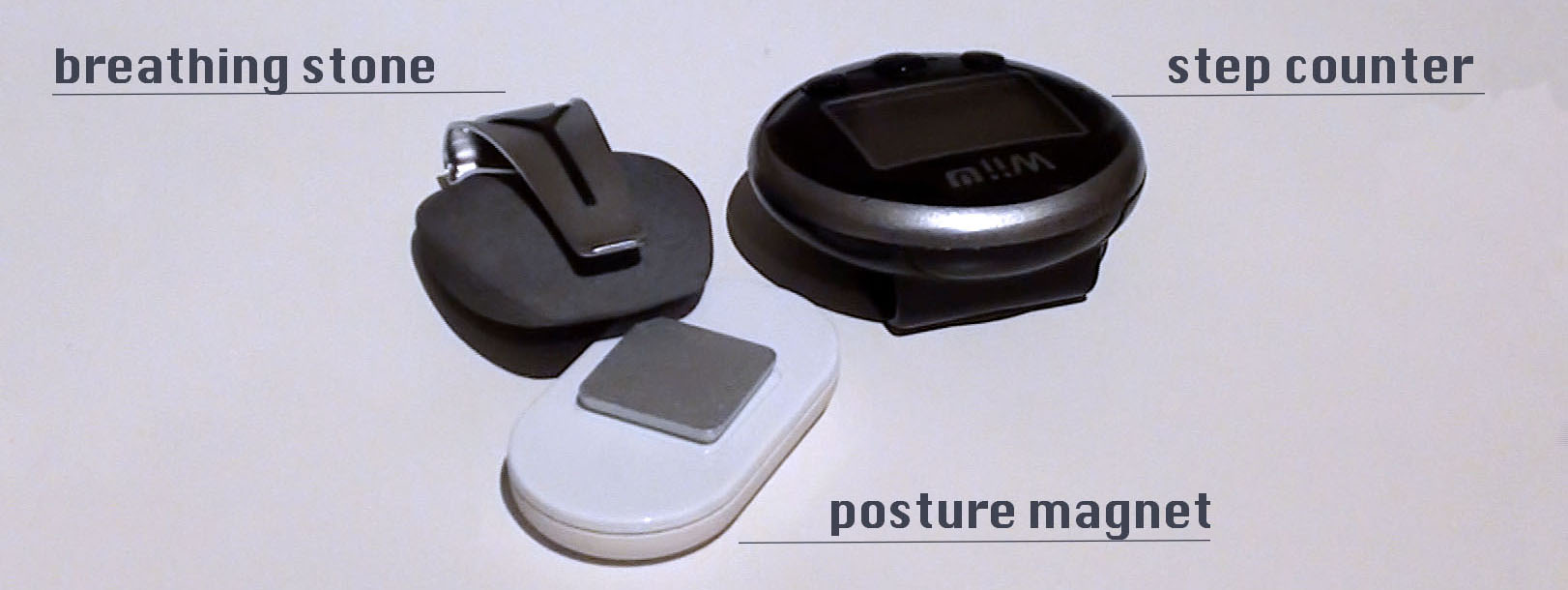}
\vspace{0em}
\caption{Scott suggests to combine these three different trackers.}\vspace{0em}
\label{fig:Scott_combination}
\end{figure}

When Scott showed us his family album he reflected:\textit{``You know at some point in life you lose your previous self and you're not yourself anymore, you're not `me' but `us' so children change everything.''} Along with the speculative combination he suggested a data album for all the members of his family. This demonstrates that Scott sees the data of the different members as a unity, hence represented in a single book. As his life has extended from the individual to the family, also the tracking has adapted and made Scott come up with new ways to track the entire family. Scott proposed that a family album could combine pictures and datasets of all the members of his family. Comparability and the ability to combine trackers and data were very important elements for him. On a personal level, he envisioned his belt as a tracking device, which would measure his posture, breathing, activity and pulse (Figure \ref{fig:Scott_combination}), or a watch that would measure his emotions and mood. These are objects that travel with him during any kind of activity.

\textit{Olivia} keeps \textit{``quite an excessive track of [her] tasks.''} She creates a list of the tasks she plans to accomplish daily (Figure \ref{fig:figure1}c). The tasks vary from very simple routine tasks, such as washing the dishes, to an important meeting at work. At the end of every day Olivia goes back to the list and deletes the accomplished tasks. The ones that have not yet been done are transferred to the next day. Thus, a task may be completed in some case even months later. Olivia also uses physical notebooks for tracking tasks. When we visited her home, Olivia showed us a shelf full of notebooks, in which most of the tasks were erased (Figure \ref{fig:figure1}d). In some cases, Olivia will go back to the notebooks after years to re-organize and she will rip off pages with tasks she does not want to remember or find important.

While in Olivia's case several personal objects were discussed, it was hard for her to detach artifacts from their origins. She came up with two main data-object combinations. The first one was a set of different color \textit{Lego} bricks, with which she could build models to represent and reflect on her data. \textit{``If I had Legos with colors and I would start making this funny platform. Like you sleep well and then you pile it up. Or make weeks out of it.''} The second was a cigarette case that would track the amount of cigarettes smoked per day. When she would try to exceed a preset limit of daily cigarettes, the case would make a disturbing sound when opened.

\begin{table*}[t!]
\centering
\small
\begin{tabular}{ p{0.15cm} | p{3cm} p{3cm} p{4cm} p{6cm} } 
 \textbf{} & \textbf{Tracking} & \textbf{Devices} & \textbf{Personal Objects} & \textbf{Data-Object Combinations} \\
 \hline
 \rotatebox[origin=c]{90}{\textbf{Hans}} & Tracks exercising in hours and kilometers, amount of weights at the gym, diet and body weight. & A smartwatch, phone applications and self-made digital spreadsheets. & Football trophies, \textit{Mad} Comic books, a snowboard bag, travel tags, a snowboard, boots, a helmet, a beanie (wool hat), winter pants, a winter jacket, a broken LP player, LP records, skiing maps. & Boots tracking location and heart rate, a skiing map showing location and skiing routes in real-time, a hat tracking speed and changing color accordingly, a book and bed sheets combo tracking sleep, earrings tracking sleep, a data LP record playing back sleep voices. \\
\hline
 \rotatebox[origin=c]{90}{\textbf{Simon}} & Tracks walking and jogging distances, spent calories, gym times and results, and different beer drank. & Phone applications, a notebook and self-made digital spreadsheets. & Soft toy duck (\textit{Puuppanen}), surfing boards, headphones, a guitar, a watch, keys and a frying pan. & Keys tracking heart rate and distance, a frying pan tracking calories and displaying it on the pan, a guitar tracking sleep and playing it back in the morning, \textit{Puuppanen} (soft toy) watching and recording sleep, a watch recording sleep while placed on the bedside table. \\ 
 \hline 
 \rotatebox[origin=c]{90}{\textbf{Max}} & Tracks work activities and hours, biking, house consuming and car driving. & Phone applications, self-made digital spreadsheets, car computer and notes. & Toy collection, a yellow collection car (\textit{Ferrari}), jeans, hand crafted cups, a cheese slicer, a popular design chair, a colorful coffee mug, a platter for keys, car and home keys, office keys, coffee table books, magazine collections and three bicycles. & A magazine providing local cycling route suggestions and displaying personal data of routes/record times, a \textit{Strava} cup visualizing cycling performance and switching on/off by putting it in a pocket, a bike lock tracking cycling performance and a key switching on/off, a piece of cloth displaying how long it lasts.  \\ 
 \hline 
 \rotatebox[origin=c]{90}{\textbf{Anna}} & Tracks movies, menstruation cycle, schedules and students' attendances to dance classes. & Phone notes, a phone application and notebooks. & A children's toolkit, childhood photos, a book, a photo book from studies, LP records, an LP player, a broken mug, \textit{Gilmore girls} quotes, CD albums, posters from London, a guitar. & A CD/LP soundtrack tracking watched movies (including one song from each seen movie), a movie mug for manually logging seen movies, a hammer and nails for recording menstruation cycle and pain, a toolkit for registering student attendances in a class, a hammer for the same purpose, a guitar melody for registering the beginning and the end of a menstruation cycle, a photo album showing important eras/periods by combining songs and videos in one book, a \textit{Spotify} playlist showing personal pictures while certain songs are playing. \\ 
 \hline 
 \rotatebox[origin=c]{90}{\textbf{Scott}} & Tracks activity (steps), body composition and temperature, posture, breathing rhythm, home electricity usage, family expenses, location, time use, living patterns (e.g., opening of doors) and more. Used to track sleep quality. & Several tracking devices: \textit{Pebble} Watch, \textit{Lumo Lift}, \textit{Spire} stone, \textit{Withings} Body Cardio, \textit{Withings} Temp, \textit{Moves} App, \textit{Google} Location and Calendar and Sheets, \textit{ScanLink OBDII + Torque}, \textit{WiiFit} Meter, \textit{Neur.io}, \textit{RescueTime}, \textit{SmartThings} hub, \textit{ThingSee} One, \textit{Exist}, \textit{Todoist} and self-made web forms. & A lion soft toy, home audio system, all the tracking devices, a children's baby book. & A Smart-watch tracking emotions and mood changes, a computer tracking the car performance, a wearable belt tracking posture, breath, steps and pulse, a family data book combining and showing different family members' data. \\
 \hline 
 \rotatebox[origin=c]{90}{\textbf{Olivia}} & Tracks schedules, tasks and sleep & Notebooks, lists on the phone and digital spreadsheets. & Cross stitching textile, notebooks, grandma's quilt, grandma's ring, rings, a small \textit{Swiss} army knife, a cigarette case, two mini dolls, old pictures, postcards, a favorite movie DVD, a \textit{Lego} figure and ceramic pieces. & \textit{Lego} bricks for tracking and visualizing sleep patterns, a smoke case tracking amount of cigarettes smoked. \\
 \hline 
\end{tabular}
\vspace{0em}
\caption{In this table we present our participants' choices in terms of tracking devices, ways to represent data, their most cherished personal objects as well as the data-object combinations they envisioned}
\vspace{-1em}
\label{tab:Table2}
\end{table*}

We have seen that our participants found many ways to interweave trackers that capture different types of data. Some of them have used different trackers that capture the same type of data for the sake of exploring the accuracy of the datasets, while others have come up with their own ways of tracking, representing and editing data (e.g., notebooks and digital spreadsheets). As the existing tracking systems allow only for certain visualizations, our participants strive to represent their personal data in ways that are appealing to them, both on a practical and aesthetic level. They have chosen their own ways of visualizing data from a wide range of options and, in some cases, invented their own mixed methods in organizing and representing data. In the following sections we discuss how these practices were reflected on their envisioned data-object combinations. 

\subsection{Data-Object Combinations}
We invited our participants to speculate on data-object combinations with a technique similar to one developed in visual anthropology \cite{pink2008visual}, in which household objects allow people to tell stories about the past, present and future. We believe that the combinations that emerged from our research are both inspiring and useful in raising awareness about the challenges of designing for future data-objects even if they were not casted as commercial products.

In Table \ref{tab:Table2}, we present all the data-object ideas (30) that came up during our study. From those, four were excluded as they were not falling under the category of self-tracking (e.g., Anna was capturing student attendance, Scott was tracking his car). 

\subsubsection{Social Sharing}
Out of the 26 combinations we analyzed, 12 ideas involved directly or indirectly another person in self-tracking. Most of the objects people own are inherently social as they connect to people, places and events in different ways. One of Simon's ideas was a frying pan that measures the calories of the meal he cooks. However, as Simon lives with his roommates, there will be times when other people will share the same pan and thus relate to the calorie measure. Max's Strava cup that would show \textit{who is the fastest} or \textit{who was elevating the most} through a change of color was placed next to other bikers' cups. This would make cyclists' performance visible to each other, while they stop for a coffee break. Also, Hans's beanie has a social aspect to it. The beanie shows the speed of the skier by changing color. Obviously, this does not make any difference to the skier since they cannot see the beanie themselves. The purpose of the color is to make other people aware of the speed of the person who is wearing it. Olivia's idea of a cigarette case that makes disturbing sounds once you have exceeded a personal limit of cigarettes per day, most likely will have an influence on other people too. Also, Hans's bed sheet tracking devices was proposed without thought of someone else sleeping with him on the same bed.   

\subsubsection{Contextual Ambiguity}
During the speculative activity, the participants proposed 13 data-objects that were completely disconnected from a tracking activity, suggesting some form of ambiguity in the interaction with the potential data systems. For instance, Simon's soft toy that would record his voice during sleep and in this way track his sleep data, represents an object that is completely detached from the type of data it tracks, but at the same time it is a very important object for the person who owns it. Anna's guitar that allows her to associate sounds to pain she feels during her menstruation cycle is another example of a disconnect between dataset and context. A more subtle example is the Strava cup. Max imagined an object that partly relates to the activity of cycling; it is used after cycling in a context where you can share your data with others. In that case, the actual cup has little to do with the cycling experience. This is in contrast to the bike lock that Max suggested, and Hans's skiing boots that are directly connected with the activity.

\subsubsection{Interacting with the Body}
17 out of the 26 imagined data-objects were both tracking devices and representations of data. The rest of the objects were trackers only, most often placed close to the body to ensure tracking is switched on. Ten out of the 26 data-objects were imagined in direct contact with the body. In the case of skiing boots, tracking is activated when the shoelaces are pulled. Hence, the data-object is switched on from the moment it is worn and set for the activity. Hans's earring, bed sheet and Simon's watch work with the same concept. From the moment the person wears the data-object the tracking begins, and it ends when the object is taken off. 

However, 14 out of the 17 combinations show data-objects that are not in direct contact with the body and act both as switches for tracking and representation. For instance,  Anna's hammer that would help her track menstruation pain, Simon's guitar that would record his sleep and play it back to him when he wakes up, Hans's skiing map that would  display his skiing routes and elevation, Olivia's Lego bricks through which she would represent her sleep and, finally, Simon's frying pan that would show the calories of the meal.

\section{Discussion}

Designing for personal data is an emerging field of research and a topic of growing interest for industry. An experience-centric approach \cite{McCarthy:2004:TE:1015530.1015549} to tracking pushes designers to think beyond numerical values and embrace the complexity of capturing and representing contextually relevant data. While data physicalization models \cite{Jansen:2015:OCD:2702123.2702180} allow an exploratory engagement with data, but detached from where and how the data was captured, data-objects acknowledge both the contextually relevant data and the physicality of objects. Our research demonstrates how individuals who employ self-tracking invent their own ways to track and represent data despite the plethora of tracking systems that currently exist in the market. Our findings highlight that personal preferences play an important role in the choices of how to self-organize data. As Pousman et al. \cite{pousman2007casual} argue, a system that visualizes data can be improved if we understand the idiosyncratic and private (and often unspoken) lives of people. However, it is a challenge for industry to shift from one-size-fits-all to designing devices for individuals. 

Looking into the private allowed us to take an experience-centric approach, after first studying the current relationships people have with their objects, we could explore how those were reflected on data-object combinations. The personal objects have their own history, which in most of the cases informed the design ideas. For instance, the experience that one participant already had with using a pair of worn-out jeans, was reflected on the data-object idea. Our research showed that it was easier for the participants to imagine data-objects that could both represent and track data, when those objects were not worn on the body. Based on the examples of the jeans and the beanie -- the only two ideas thought both as worn representations and trackers -- we can speculate that it only makes sense for people to represent their personal data on the body if the data is also relevant for others.  

\subsection{Reflecting Data Experiences}
According to Lupton \cite{lupton2016personal} people have always used objects as technologies, but with technologies becoming smaller and easier to use it is less obvious where the body ends and were the technology begins. The participants conceived the data-objects as inseparable from their everyday routines. This supports Elsden et al.'s argument that for personal informatics tools to be meaningful, the data needs to resonate with people's lived experiences \cite{Elsden2016AInformatics}. However, people's lived experiences are so complex and unique that a designer can only capture a small part of them when designing for data in context. 

Our research suggests that data-objects can function without direct coupling to the tracking activity and that ambiguous objects may open a space for richer interpretations and reflections on data. As suggested by Gaver et al. \cite{gaver2003ambiguity} artifacts seen in a context different from their origin may acquire different meanings. Data physicalizations \cite{Jansen:2015:OCD:2702123.2702180} achieve that ambiguity through different material assemblages, which enable rich self-reflections. In data-objects this ambiguity can perhaps be achieved by unexpected couplings of data and objects. 

The soft toy that watches your sleep and the guitar playing a lullaby resembling sleep data are examples that bank on metaphors to connect to peoples' childhood and memories. We see an advantage in metaphors that relate to everyday objects shared among people and rooted in their culture as for instance, a doll that will watch your sleep. We thus invite designers of tracking devices to embrace these type of metaphors with the purpose to bring poetry and relatedness into daily tracking. 

\subsection{Social Sharing}
Data is inherently social, with a range of social rationales potentially being connected to it. In our research, the participants wanted to combine their datasets with those of other people, like their family members (Scott) or their flat mates (Hans). We observed this both in their current data practices, for example Scott's online tracking system for his family's expenses, and in their ideas for envisioning data-objects, such as Max's suggestion of a magazine that compares his performance with that of other cyclists. This is not an odd observation if we think that people tend to understand themselves in relation to others. Oneself is reflected through the eyes of others, thus, the word \textit{me} is both general and individual \cite{Cooley1902,Goffman1982}. 

The quantified-self forum\footnote{http://quantifiedself.com/} is a clear example of people discussing their data with others on a shared platform. There may be a dichotomy, though. While Scott imagines the different family data sets as a unity, Hans and Max see their data sets in comparison with others. The way people see their data connecting to others' is a challenge for how current tracking systems are built. Prior work has investigated opportunities for family-based tracking using coasters to represent data relevant to the family (e.g., \cite{Singhal2017Time-Turner:Home}). That work suggests that we should acknowledge that personal data does not always focus solely on the individual. Instead, it should be thought of as a social entity that can be shared among people for different purposes. When thinking of sharing of data-objects we should also reflect on how other people might interfere indirectly with individual tracking through daily interactions (e.g., bed sheets). For instance, starting from the family and expanding towards larger communities of people that work towards the same goal. We invite designers to explore what shared goals people might have with practical everyday tasks, such as finances of household, to more exploratory ones, such as exercise competition between roommates. We believe that this turn towards self-exploration through tracking requires more playful systems that, apart from motivating behavioral change, may promote collaboration and experimentation with other people through a single shared artifact that is aware of the activity of others around us. 

Our work agrees with Rooksby et al.'s \cite{Rooksby:2014:PTL:2556288.2557039} argument that tracking needs to be done on the basis of lived experiences. We extend their rationale by proposing that it is important to think of the poetics and complexity of everyday life that might not always be directly linked to context. Objects have the power of becoming meaningful to people, they are private but also social. They can be brought into the foreground and the background of peoples' lives at the same time. In this way designers may enrich the current relationships between lived experiences and data.

\subsection{Reflections on Methodology}
 Reflecting on the way we invited people to imagine data-object combinations, one may discuss, if we set an expectation for the participant to come up with something original that made them hesitate to express their thinking. For instance, Anna was hesitant to suggest speculative combinations. In the beginning of the object theater exercise she made statements such as, \textit{``I am blank''} or when asked if there is anything that comes into her mind she replied, \textit{``Nothing that creative really.''} We countered this by challenging her to propose ideas that were unconventional focusing on one object at a time. For instance, \textit{``which of these objects (pointing to the objects assembly) would you track your sleep data with?''} Overall, as an invitation to play with the object assemblies, the ideas were created in mutual improvisation. As suggested in other work with object field studies \cite{ryoppy2018object}, the researchers' participation is crucial in inviting the participant to shake free of limitations. A way to do that is to suggest `wild' ideas and challenge people to decide whether the combination makes sense and what interaction styles it might include. 

Considering that it was hard for people to transition from presenting the objects to imagining data-object combinations, we propose for future development of the method that researchers should use warm up exercises. For example, by inviting people to make free associations to their personal objects or even by beginning the warm up exercise with objects the researchers bring with them, so objects do not have sentimental meaning to people.

\section{Conclusion}
In this paper, we present an experience-centric point of view on data-objects. Based on speculative ideas of data-object combinations created in object theater activities with six participants, we identified three aspects that can enhance the design of data-objects: social sharing, contextual ambiguity and interaction with the body. We suggest that when designing for self-tracking, designers should consider people's idiosyncratic characteristics. New designs could be based on metaphors rooted in the history of objects that fit the living context both on a personal and a collective level. In the future, we plan to develop design prototypes to further explore the concept of data-objects that encapsulate the poetics of everyday life in their use and properties.

\section{Acknowledgments}
We thank the participants of our study who allowed us into their homes and shared with us their imaginative thinking.

\balance{}

\bibliographystyle{SIGCHI-Reference-Format}
\bibliography{data-objects}


\begin{thebibliography}{00}


\ifx \showCODEN    \undefined \def \showCODEN     #1{\unskip}     \fi
\ifx \showDOI      \undefined \def \showDOI       #1{{\tt DOI:}\penalty0{#1}\ }
  \fi
\ifx \showISBNx    \undefined \def \showISBNx     #1{\unskip}     \fi
\ifx \showISBNxiii \undefined \def \showISBNxiii  #1{\unskip}     \fi
\ifx \showISSN     \undefined \def \showISSN      #1{\unskip}     \fi
\ifx \showLCCN     \undefined \def \showLCCN      #1{\unskip}     \fi
\ifx \shownote     \undefined \def \shownote      #1{#1}          \fi
\ifx \showarticletitle \undefined \def \showarticletitle #1{#1}   \fi
\ifx \showURL      \undefined \def \showURL       #1{#1}          \fi

\bibitem{Ayobi:2018:FMS:3173574.3173602}
{Amid Ayobi}, {Tobias Sonne}, {Paul Marshall}, {and} {Anna~L. Cox}. 2018.
\newblock \showarticletitle{Flexible and Mindful Self-Tracking: Design
  Implications from Paper Bullet Journals}. In {\em Proceedings of the 2018 CHI
  Conference on Human Factors in Computing Systems} {\em (CHI '18)}. ACM, New
  York, NY, USA, Article 28, 14 pages.
\newblock
\showISBNx{978-1-4503-5620-6}
\showDOI{%
\url{http://dx.doi.org/10.1145/3173574.3173602}}


\bibitem{Baker}
{Rene Baker}. 2018.
\newblock Specialist in puppet and object theatre.
\newblock   (2018).
\newblock
\showURL{%
\url{https://renebaker.org/}}


\bibitem{Barrass2012}
{Stephen Barrass}. 2012.
\newblock \showarticletitle{{Digital fabrication of acoustic sonifications}}.
\newblock {\em AES: Journal of the Audio Engineering Society\/} {60}, 9 (2012),
  709--715.
\newblock
\showISSN{15494950}


\bibitem{44b44fd5eaee419ca9729b85acb875ef}
{Jacob Buur} {and} {Preben Friis}. 2015.
\newblock \showarticletitle{Object Theatre in Design Education}. In {\em Nordes
  2015}.
\newblock


\bibitem{Buur2018Physicalizations}
{Jacob Buur}, {Sara~Said Mosleh}, {and} {Christina Fyhn}. 2018.
\newblock \showarticletitle{{Physicalizations of Big Data in Ethnographic
  Context}}.
\newblock {\em Ethnographic Praxis in Industry Conference Proceedings\/}
  {2018}, 1 (2018), 86--103.
\newblock
\showDOI{%
\url{http://dx.doi.org/10.1111/1559-8918.2018.01198}}


\bibitem{Choe:2014:UQP:2556288.2557372}
{Eun~Kyoung Choe}, {Nicole~B. Lee}, {Bongshin Lee}, {Wanda Pratt}, {and}
  {Julie~A. Kientz}. 2014.
\newblock \showarticletitle{Understanding Quantified-selfers' Practices in
  Collecting and Exploring Personal Data}. In {\em Proceedings of the SIGCHI
  Conference on Human Factors in Computing Systems} {\em (CHI '14)}. ACM, New
  York, NY, USA, 1143--1152.
\newblock
\showISBNx{978-1-4503-2473-1}
\showDOI{%
\url{http://dx.doi.org/10.1145/2556288.2557372}}


\bibitem{Cooley1902}
{Charles~Horton Cooley}. 1902.
\newblock \showarticletitle{{The Looking-Glass Self}}.
\newblock  (1902), 1--2.
\newblock


\bibitem{DeLeon2005}
{Jason~Patrick de Leon} {and} {Jeffrey~H. Cohen}. 2005.
\newblock \showarticletitle{{Object and Walking Probes in Ethnographic
  Interviewing}}.
\newblock {\em Field Methods\/} {17}, 2 (2005), 200--204.
\newblock
\showISSN{1525822X}
\showDOI{%
\url{http://dx.doi.org/10.1177/1525822X05274733}}


\bibitem{Elsden2017DesigningInformatics}
{Chris Elsden}, {Abigail~C. Durrant}, {David Chatting}, {and} {David~S. Kirk}.
  2017.
\newblock \showarticletitle{{Designing Documentary Informatics}}.
\newblock  (2017), 649--661.
\newblock
\showISBNx{9781450349222}
\showDOI{%
\url{http://dx.doi.org/10.1145/3064663.3064714}}


\bibitem{Elsden2016AInformatics}
{Chris Elsden}, {David~S. Kirk}, {and} {Abigail~C. Durrant}. 2016.
\newblock \showarticletitle{{A Quantified Past: Toward Design for Remembering
  With Personal Informatics}}.
\newblock {\em Human-Computer Interaction\/} {31}, 6 (2016), 518--557.
\newblock
\showISSN{07370024}
\showDOI{%
\url{http://dx.doi.org/10.1080/07370024.2015.1093422}}


\bibitem{Frick}
{Laurie Frick}. 2011.
\newblock Sleep Patterns.
\newblock   (2011).
\newblock
\showURL{%
Retrieved September 9, 2019 from \url{https://vimeo.com/21852158}}


\bibitem{Gaver:2012:AP:2212877.2212889}
{Bill Gaver} {and} {John Bowers}. 2012.
\newblock \showarticletitle{Annotated Portfolios}.
\newblock {\em Interactions\/} {19}, 4 (July 2012), 40--49.
\newblock
\showISSN{1072-5520}
\showDOI{%
\url{http://dx.doi.org/10.1145/2212877.2212889}}


\bibitem{gaver2003ambiguity}
{William~W Gaver}, {Jacob Beaver}, {and} {Steve Benford}. 2003.
\newblock \showarticletitle{Ambiguity as a resource for design}. In {\em
  Proceedings of the SIGCHI conference on Human factors in computing systems}.
  ACM, 233--240.
\newblock


\bibitem{Ghosh}
{Choiti Ghosh}. 2016.
\newblock Ordinary Objects in Theater: Breaking Norms | Choiti Ghosh |
  TEDxMDAE.
\newblock   (2016).
\newblock
\showURL{%
\url{https://www.youtube.com/watch?v=FDLAx3dd_LU}}


\bibitem{Goffman1982}
{Erving Goffman}. 1982.
\newblock \showarticletitle{{The presentation of Self in Everyday Life}}.
\newblock {\em the Presentation of Self in Everyday Life\/} (1982), 1--10.
\newblock


\bibitem{Jansen:2015:OCD:2702123.2702180}
{Yvonne Jansen}, {Pierre Dragicevic}, {Petra Isenberg}, {Jason Alexander},
  {Abhijit Karnik}, {Johan Kildal}, {Sriram Subramanian}, {and} {Kasper
  Hornb{\ae}k}. 2015.
\newblock \showarticletitle{Opportunities and Challenges for Data
  Physicalization}. In {\em Proceedings of the 33rd Annual ACM Conference on
  Human Factors in Computing Systems} {\em (CHI '15)}. ACM, New York, NY, USA,
  3227--3236.
\newblock
\showISBNx{978-1-4503-3145-6}
\showDOI{%
\url{http://dx.doi.org/10.1145/2702123.2702180}}


\bibitem{Karyda:2017:CCI:3098279.3119927}
{Maria Karyda}. 2017.
\newblock \showarticletitle{Crafting Collocated Interactions: Exploring
  Physical Representations of Personal Data}. In {\em Proceedings of the 19th
  International Conference on Human-Computer Interaction with Mobile Devices
  and Services} {\em (MobileHCI '17)}. ACM, New York, NY, USA, Article 74, 4
  pages.
\newblock
\showISBNx{978-1-4503-5075-4}
\showDOI{%
\url{http://dx.doi.org/10.1145/3098279.3119927}}


\bibitem{Klepp2014}
{Ingun~Grimstad Klepp} {and} {Mari Bjerck}. 2014.
\newblock \showarticletitle{{A methodological approach to the materiality of
  clothing: Wardrobe studies}}.
\newblock {\em International Journal of Social Research Methodology\/} {17}, 4
  (2014), 373--386.
\newblock
\showISSN{14645300}
\showDOI{%
\url{http://dx.doi.org/10.1080/13645579.2012.737148}}


\bibitem{latour1987science}
{Bruno Latour}. 1987.
\newblock {\em Science in action: How to follow scientists and engineers
  through society}.
\newblock Harvard university press.
\newblock


\bibitem{li2010stage}
{Ian Li}, {Anind Dey}, {and} {Jodi Forlizzi}. 2010.
\newblock \showarticletitle{A stage-based model of personal informatics
  systems}. In {\em Proceedings of the SIGCHI conference on human factors in
  computing systems}. ACM, 557--566.
\newblock


\bibitem{Li:2011:UMD:2030112.2030166}
{Ian Li}, {Anind~K. Dey}, {and} {Jodi Forlizzi}. 2011.
\newblock \showarticletitle{Understanding My Data, Myself: Supporting
  Self-reflection with Ubicomp Technologies}. In {\em Proceedings of the 13th
  International Conference on Ubiquitous Computing} {\em (UbiComp '11)}. ACM,
  New York, NY, USA, 405--414.
\newblock
\showISBNx{978-1-4503-0630-0}
\showDOI{%
\url{http://dx.doi.org/10.1145/2030112.2030166}}


\bibitem{Lupton:2014:SCT:2686612.2686623}
{Deborah Lupton}. 2014.
\newblock \showarticletitle{Self-tracking Cultures: Towards a Sociology of
  Personal Informatics}. In {\em Proceedings of the 26th Australian
  Computer-Human Interaction Conference on Designing Futures: The Future of
  Design} {\em (OzCHI '14)}. ACM, New York, NY, USA, 77--86.
\newblock
\showISBNx{978-1-4503-0653-9}
\showDOI{%
\url{http://dx.doi.org/10.1145/2686612.2686623}}


\bibitem{lupton2016personal}
{Deborah Lupton}. 2016.
\newblock \showarticletitle{Personal data practices in the age of lively data}.
\newblock {\em Digital sociologies\/} (2016), 335--50.
\newblock


\bibitem{margolies2016props}
{Eleanor Margolies}. 2016.
\newblock {\em Props}.
\newblock Macmillan International Higher Education.
\newblock


\bibitem{McCarthy:2004:TE:1015530.1015549}
{John McCarthy} {and} {Peter Wright}. 2004.
\newblock \showarticletitle{Technology As Experience}.
\newblock {\em Interactions\/} {11}, 5 (Sept. 2004), 42--43.
\newblock
\showISSN{1072-5520}
\showDOI{%
\url{http://dx.doi.org/10.1145/1015530.1015549}}


\bibitem{miller2008comfort}
{Daniel Miller}. 2008.
\newblock {\em The comfort of things}.
\newblock Polity.
\newblock


\bibitem{moere2008beyond}
{Andrew~Vande Moere}. 2008.
\newblock \showarticletitle{Beyond the tyranny of the pixel: Exploring the
  physicality of information visualization}. In {\em 2008 12th International
  Conference Information Visualisation}. IEEE, 469--474.
\newblock


\bibitem{Moller:2018:PAA:3173574.3174133}
{Trine M{\o}ller}. 2018.
\newblock \showarticletitle{Presenting The Accessory Approach: A Start-up's
  Journey Towards Designing An Engaging Fall Detection Device}. In {\em
  Proceedings of the 2018 CHI Conference on Human Factors in Computing Systems}
  {\em (CHI '18)}. ACM, New York, NY, USA, Article 559, 10 pages.
\newblock
\showISBNx{978-1-4503-5620-6}
\showDOI{%
\url{http://dx.doi.org/10.1145/3173574.3174133}}


\bibitem{mortier2014human}
{Richard Mortier}, {Hamed Haddadi}, {Tristan Henderson}, {Derek McAuley}, {and}
  {Jon Crowcroft}. 2014.
\newblock \showarticletitle{Human-data interaction: The human face of the
  data-driven society}.
\newblock {\em Available at SSRN 2508051\/} (2014).
\newblock


\bibitem{myatt2012frozen}
{Sean Myatt} {and} {Daniel Watt}. 2012.
\newblock \showarticletitle{From Frozen Sponges to Plastic Bags: Object
  Theatre-A Developing Network}.
\newblock {\em Puppet Notebook\/}  {22} (2012), 19--22.
\newblock


\bibitem{Nissen:2015:DDF:2702123.2702245}
{Bettina Nissen} {and} {John Bowers}. 2015.
\newblock \showarticletitle{Data-Things: Digital Fabrication Situated Within
  Participatory Data Translation Activities}. In {\em Proceedings of the 33rd
  Annual ACM Conference on Human Factors in Computing Systems} {\em (CHI '15)}.
  ACM, New York, NY, USA, 2467--2476.
\newblock
\showISBNx{978-1-4503-3145-6}
\showDOI{%
\url{http://dx.doi.org/10.1145/2702123.2702245}}


\bibitem{nordstrom2013object}
{Susan~Naomi Nordstrom}. 2013.
\newblock \showarticletitle{Object-interviews: Folding, unfolding, and
  refolding perceptions of objects}.
\newblock {\em International Journal of Qualitative Methods\/} {12}, 1 (2013),
  237--257.
\newblock


\bibitem{pink2008visual}
{Sarah Pink}. 2008.
\newblock \showarticletitle{Visual anthropology}.
\newblock  (2008).
\newblock


\bibitem{pousman2007casual}
{Zachary Pousman}, {John Stasko}, {and} {Michael Mateas}. 2007.
\newblock \showarticletitle{Casual information visualization: Depictions of
  data in everyday life}.
\newblock {\em IEEE transactions on visualization and computer graphics\/}
  {13}, 6 (2007), 1145--1152.
\newblock


\bibitem{Rezaeian:2014:DTD:2636240.2636869}
{Alireza Rezaeian} {and} {Jared Donovan}. 2014.
\newblock \showarticletitle{Design of a Tangible Data Visualization}. In {\em
  Proceedings of the 7th International Symposium on Visual Information
  Communication and Interaction} {\em (VINCI '14)}. ACM, New York, NY, USA,
  Article 232, 4 pages.
\newblock
\showISBNx{978-1-4503-2765-7}
\showDOI{%
\url{http://dx.doi.org/10.1145/2636240.2636869}}


\bibitem{Rooksby:2014:PTL:2556288.2557039}
{John Rooksby}, {Mattias Rost}, {Alistair Morrison}, {and} {Matthew Chalmers}.
  2014.
\newblock \showarticletitle{Personal Tracking As Lived Informatics}. In {\em
  Proceedings of the SIGCHI Conference on Human Factors in Computing Systems}
  {\em (CHI '14)}. ACM, New York, NY, USA, 1163--1172.
\newblock
\showISBNx{978-1-4503-2473-1}
\showDOI{%
\url{http://dx.doi.org/10.1145/2556288.2557039}}


\bibitem{ryoppy2018object}
{Merja Ry{\"o}ppy}, {Sofus~Bach Poulsen}, {Pavels Konstantinovs}, {and} {Salu
  Ylirisku}. 2018.
\newblock \showarticletitle{Object theatre in field studies}. In {\em 5th
  Participatory Innovation ConferenceParticipatory Innovation Conference}.
  286--293.
\newblock


\bibitem{ryoppy2017exploring}
{Merja Ry{\"o}ppy}, {Salu Ylirisku}, {and} {Eva Knutz}. 2017.
\newblock \showarticletitle{Exploring Power with Object Theatre}. In {\em 7th
  Nordic Design Research ConferenceNordic Design Research Conference}.
\newblock


\bibitem{Singhal2017Time-Turner:Home}
{Samarth Singhal}, {William Odom}, {Lyn Bartram}, {and} {Carman Neustaedter}.
  2017.
\newblock \showarticletitle{{Time-Turner: Data Engagement Through Everyday
  Objects in the Home}}.
\newblock {\em Proceedings of the 2017 ACM Conference Companion Publication on
  Designing Interactive Systems\/} (2017), 72--78.
\newblock
\showISBNx{978-1-4503-4991-8}
\showDOI{%
\url{http://dx.doi.org/10.1145/3064857.3079122}}


\bibitem{Sosa2018}
{R. Sosa}, {V. Gerrard}, {A. Esparza}, {R. Torres}, {and} {R. Napper}. 2018.
\newblock \showarticletitle{{Data objects: Design principles for data
  physicalisation}}.
\newblock {\em Proceedings of International Design Conference, DESIGN\/}  {4}
  (2018), 1685--1696.
\newblock
\showISBNx{9789537738594}
\showISSN{18479073}
\showDOI{%
\url{http://dx.doi.org/10.21278/idc.2018.0125}}


\bibitem{taylor2015data}
{Alex~S Taylor}, {Si{\^a}n Lindley}, {Tim Regan}, {David Sweeney}, {Vasillis
  Vlachokyriakos}, {Lillie Grainger}, {and} {Jessica Lingel}. 2015.
\newblock \showarticletitle{Data-in-place: Thinking through the relations
  between data and community}. In {\em Proceedings of the 33rd Annual ACM
  Conference on Human Factors in Computing Systems}. ACM, 2863--2872.
\newblock


\bibitem{williams2015anxious}
{Kaiton Williams}. 2015.
\newblock \showarticletitle{An anxious alliance}. In {\em Proceedings of The
  Fifth Decennial Aarhus Conference on Critical Alternatives}. Aarhus
  University Press, 121--131.
\newblock


\bibitem{wolf2010data}
{Gary Wolf}. 2010.
\newblock \showarticletitle{The data-driven life}.
\newblock {\em The New York Times\/}  {28} (2010), 2010.
\newblock


\bibitem{woodward2016object}
{Sophie Woodward}. 2016.
\newblock \showarticletitle{Object interviews, material imaginings and
  'unsettling'methods: interdisciplinary approaches to understanding materials
  and material culture}.
\newblock {\em Qualitative Research\/} {16}, 4 (2016), 359--374.
\newblock


\bibitem{Zhu2015Data-objects:Interfaces}
{Chang~Long Zhu}, {Harshit Agrawal}, {and} {Pattie Maes}. 2015.
\newblock \showarticletitle{{Data-objects: Re-designing everyday objects as
  tactile affective interfaces}}.
\newblock {\em 2015 International Conference on Affective Computing and
  Intelligent Interaction, ACII 2015\/} (2015), 322--326.
\newblock
\showISBNx{9781479999538}
\showDOI{%
\url{http://dx.doi.org/10.1109/ACII.2015.7344590}}


\end{thebibliography}

\end{document}